\documentclass[journal = cmatex, article, traditional, titlepage, layout=twocolumn,citeautoscript]{achemso}

\usepackage{dcolumn}
\usepackage{amssymb}
\usepackage{float}
\usepackage{graphicx}
\usepackage{amsmath}
\usepackage{booktabs}
\usepackage{multirow}
\usepackage{array}
\usepackage{makecell}
\usepackage{natbib}
\usepackage{textcomp}
\usepackage{bm}
\usepackage{color}
\usepackage[utf8]{inputenc}
\usepackage[T1]{fontenc}

\definecolor{darkgreen}{rgb}{0,.6,0}

\setcitestyle{super}

\author{Andrew J. E. Rowberg}

\author{Chris G. Van de Walle}
\email{vandewalle@mrl.ucsb.edu}
\affiliation{Materials Department, University of California, Santa Barbara, California 93106-5050, United States}

\title{Hydride Conductivity in Nitride Hydrides}

\begin{document}

\begin{abstract}

Nitride hydrides are a largely unexplored class of materials with promising applications in solid-state hydrogen fuel cells.
Here, we use first-principles calculations to characterize defects and ionic mobility in Sr$_2$LiH$_2$N (SLHN), a nitride hydride with high hydride conductivity.
Calculating defect formation energies, we find that SLHN contains high concentrations of hydrogen interstitials (H$_i$). 
H$_i^-$ migrates with very low energetic barriers, which, together with its low formation energy, implies that SLHN will have excellent hydride kinetics, potentially surpassing those of other known hydride electrolytes.
Oxygen contamination is a concern, meaning that encapsulation will be critical.
By direct analogy to the La/Sr-based oxyhydrides, which have similar crystal structures, we also investigate the La-based nitride hydride La$_2$LiHN$_2$ but find that it will be significantly less 
conductive, and thus not as technologically useful.
Our findings buttress the exploration of SLHN and similar nitride hydrides for use in solid-state hydrogen fuel cells.

\end{abstract}

\maketitle

\section{Introduction}
\label{intro}

Hydrogen fuel cell technology has expanded in recent years, due to hydrogen's high gravimetric energy density,
its elemental abundance, and the growing need for renewable energy sources that do not emit carbon dioxide.\cite{schlapbach_hydrogen-storage_2001,SteeleNAT2001,crabtree2008hydrogen,IEA2019}
Current hydrogen fuel cells incorporate polymer electrolytes, typically Nafion\texttrademark, to transport hydrogen ions and drive an electric current.\cite{rikukawa2000proton,wang2011review}
However, one outstanding goal of hydrogen energy research is to 
develop solid-state fuel cells, which offer higher operating temperatures and superior stability, to which end it is necessary to replace this polymer with a solid-state electrolyte.\cite{SteeleNAT2001,jena2011materials}
Most solid-state electrolytes under investigation are proton-conducting oxides, notably BaZrO$_3$.\cite{kreuer2003proton,iwahara_zirc_1993,StambouliRSER2002}
However, recent research has also focused on metal hydrides, such as BaH$_2$,\cite{verbraeken_high_2014} and more complex mixed-anionic oxyhydrides, such as La$_{2-x}$Sr$_x$LiH$_{1+x}$O$_{3-x}$.\cite{kobayashi2016pure,kobayashi_oxyhydride_2012,bridges2006observation}
These systems differ from traditional proton conducting oxides in several ways: they intrinsically contain hydrogen; the mobile species is the hydride ion (H$^-$), rather than the proton (H$^+$); and diffusion typically proceeds via vacancy migration.\cite{rowberg_AeH2_2018,bai2018first,liu2018formation}
Mixed-anionic materials in general provide interesting scientific prospects by expanding the design space created by traditional single-anion materials.
However, virtually all of the known mixed-anionic crystals contain oxygen, and only a small fraction of those contain hydrogen.\cite{kageyama2018expanding}
Thus, further exploration of hydridic mixed-anionic crystals is warranted.

One largely unexplored class of materials are nitride hydrides (or ``hydridonitrides''\cite{aleksanyan2001combustion}), which are mixed-anionic systems containing nitrogen and hydrogen.
Researchers have known about the high hydrogen storage capacity (up to 11.5 wt\%) of the Li-N-H system for several decades,\cite{chen2002interaction} and efficient synthesis methods to generate Li$_4$HN are available.\cite{tapia2013rapid}
Alkaline-earth nitride hydrides, such as Sr$_2$NH and Ba$_2$NH, have also been synthesized,\cite{reckeweg2002alkaline,chemnitzer2005sr2n} and they have been shown to exhibit high hydride diffusion.\cite{wegner1992structure,altorfer1994Ba2NH}

Two groups have synthesized and characterized Sr$_2$LiH$_2$N (SLHN),\cite{blaschkowski2007darstellung,liu2010synthesis} though ion transport was not measured,
This material has a very similar structure to La$_{2-x}$Sr$_x$LiH$_{1+x}$O$_{3-x}$ (LSLHO), the only oxyhydride to have been demonstrated in a fuel cell.\cite{kobayashi2016pure,matsui2019ambient}
Kobayashi $et$ $al.$ noted that hydride conductivity increases as the Sr-to-La ratio increases in LSLHO;\cite{kobayashi2016pure} however, as we have previously seen, the stability of the oxyhydride decreases with increasing Sr content.\cite{rowberg_oxyhydrides_2020}
Thus, it is encouraging that the first experimental reports on the analogous nitride hydride system center on the Sr-containing compound.
The structure of SLHN also contains free space in the form of linear cavities for hydride ions to migrate through the crystal, potentially providing pathways for hydride diffusion.\cite{blaschkowski2007darstellung}
These features suggest that it should exhibit high hydride conductivity.

In this paper, we comprehensively evaluate SLHN's prospects as a solid-state hydrogen electrolyte.
We use first-principles calculations based on density functional theory (DFT)\cite{kohn_self-consistent_1965,hohenberg1964} and employing a hybrid functional,\cite{heyd_hse} which allows us to calculate the energetics for defect formation with high accuracy.
We focus on a region of chemical potentials in SLHN for which defect formation energies are positive.
We identify hydrogen interstitials and vacancies as the most common native defects in SLHN.
These defects, interstitials in particular, mediate the hydride conductivity, which we quantify through the use of nudged elastic band (NEB) calculations.\cite{henkelman_neb}
Our calculated activation barrier for hydride diffusion is lower than the measured barrier in LaHO,\cite{fukui2019characteristic} the most conductive oxyhydride known.
Among hydride conductors studied to date, SLHN is unique in that interstitial hydrogen (rather than the hydrogen vacancy) is the primary defect driving hydride migration.
We also use our methodology to assess the energetics of oxygen incorporation, highlighting the need to protect the material from oxygen.
Finally, in analogy with the La/Sr oxyhydride system, we discuss the prospects of a hypothetical La-containing material with the chemical formula La$_2$LiHN$_2$ (LLHN), which could, in principle, marry the advantages of chemical stability and high hydride conductivity.
However, we find LLHN to be both unstable and less conductive than SLHN, making it less technological interesting.
In sum, our results clearly show that SLHN is an excellent hydride conductor and one that merits further experimental exploration for use in solid-state hydride fuel cells.

\section{Methodology}
\label{method}

Our calculations are based on DFT within the generalized Kohn-Sham scheme,\cite{kohn_self-consistent_1965} as implemented in the Vienna $Ab$ $initio$ Simulation Package (VASP) \cite{kresse_vasp}.
We use the hybrid exchange-correlation functional of Heyd, Scuseria, and Ernzerhof (HSE),\cite{heyd_hse} with 25\% mixing of short-range Hartree-Fock exchange.
We apply projector augmented wave (PAW) potentials~\cite{blochl_paw1,kresse_paw2} with a plane-wave cutoff of 500 eV.
The Sr $4s^2$ $4p^6$ $5s^2$, La $5s^2$ $5p^6$ $6s^2$ $5d^1$, Li $1s^2$ $2s^1$, and N $2s^2$ $2p^3$ electrons are treated explicitly as valence.
We simulate orthorhombic unit cells for SLHN and LLHN, each containing four formula units (see Fig.~\ref{fig:struct}), using an 8$\times$4$\times$2 $k$-point grid to integrate over the Brillouin zone.
For our defect calculations, we construct 3$\times$2$\times$1 supercells to ensure a length greater than 10 \AA \:in each Cartesian direction.
We use the NEB method with climbing images to evaluate the energetic barriers associated with defect migration.\cite{henkelman_neb}
For our NEB calculations, we use the generalized gradient approximation functional of Perdew, Burke, and Ernzerhof\cite{perdew1996generalized} with three intermediate images.

\section{Results and Discussion}
\label{sec:res}

\subsection{Structure}
\label{sec:struc}

\begin{figure}
\includegraphics[width=9cm]{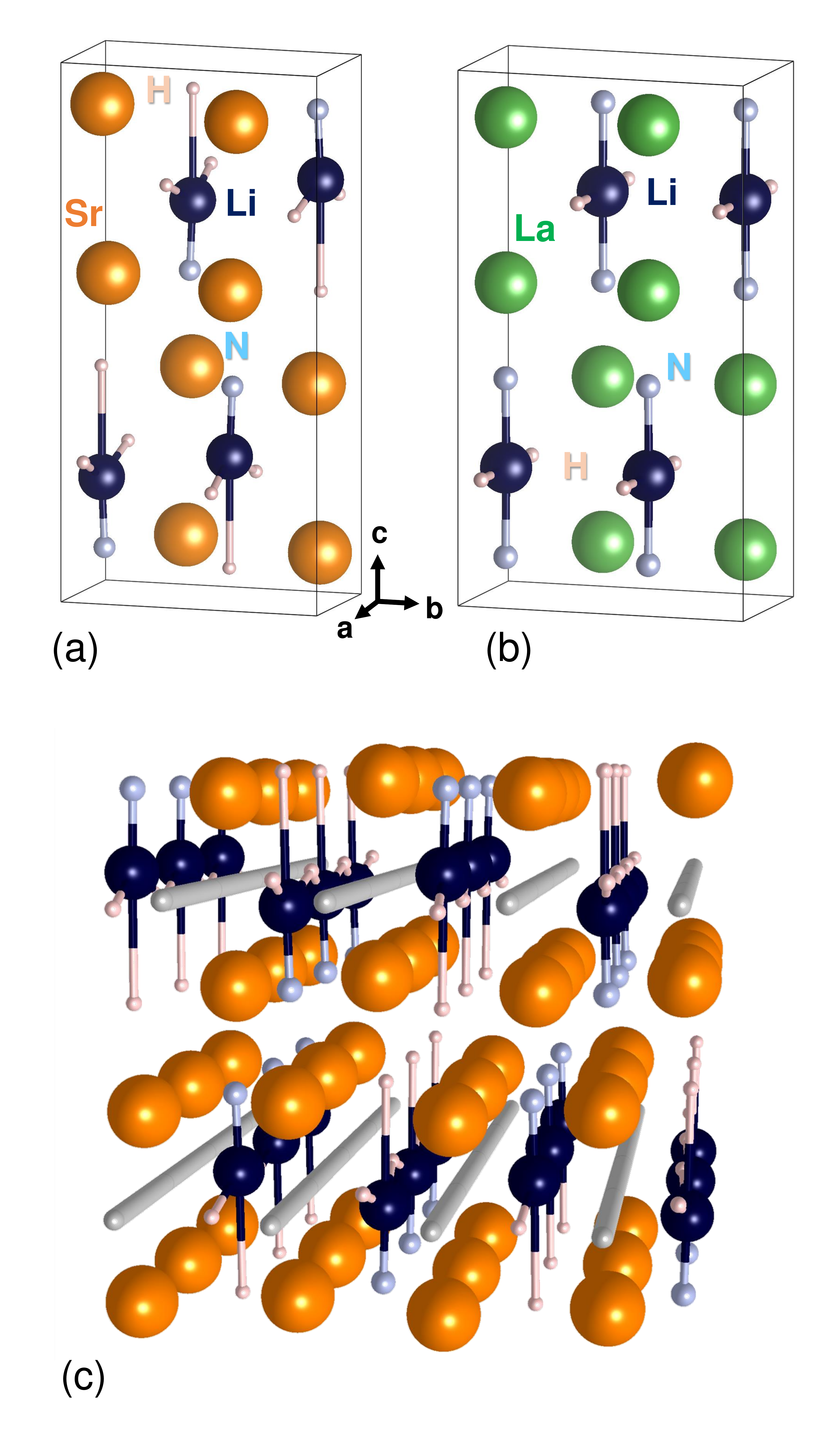}
\caption{(a) Unit cells of (a) Sr$_2$LiH$_2$N and (b) La$_2$LiHN$_2$.
(c) A 3$\times$2$\times$1 supercell of SLHN, with empty channels for hydride migration indicated with thick lines. Images generated using the VESTA 3 software.\cite{momma2011vesta}}
\label{fig:struct}
\end{figure}

Experimental studies suggested two possible structures for SLHN, both with $a$-, $b$-, and $c$-axes oriented at right angles relative to one another.
Both derive from fourfold-coordinated planar motifs of Li bonded to four anions, one N and three H, with Sr cations simply providing charge balance.
In the Li--N--3H motifs, the N anion and one of the H anions are oriented apically along the long $c$ axis relative to Li. The other two H anions could be oriented either along the $a$ or $b$ axes relative to Li.
In Ref.~\citenum{blaschkowski2007darstellung}, crystallographic analysis revealed a preference for these H anions to lie uniformly along one axis [chosen to be the $a$ axis in Fig.~\ref{fig:struct}(a)], thereby forming chains of Li--N--3H motifs connected by shared H anions along the $a$ axis [see Fig.~\ref{fig:struct}(c)].
That structure is orthorhombic, with a $Pnma$ space group.
The authors of Ref.~\citenum{liu2010synthesis}, on the other hand, proposed a tetragonal structure (space group $I4/mmm$) with $a=b$.
Their methods were not sensitive to the location of H atoms, but their structure would be consistent with the same apical H ordering, while the remaining two H atoms would be randomly oriented along the $a$ or $b$ axes relative to Li (i.e., the planar Li motifs lie either in the $ac$ or the $bc$ plane).
Structural parameters for both experimental structures are listed in Table~\ref{tab:bulk}.

For our purposes, we assume the orthorhombic $Pnma$ structure of Ref.~\citenum{blaschkowski2007darstellung} for both SLHN and the hypothetical material LLHN.
We also investigated the tetragonal structure of Ref.~\citenum{liu2010synthesis}, but we found it to have a higher energy.
We show the orthorhombic unit cells for SLHN in Fig.~\ref{fig:struct}(a) and for LLHN in Fig.~\ref{fig:struct}(b), and we list our calculated bulk parameters in Table~\ref{tab:bulk} alongside experimental parameters for SLHN.
Per unit cell, there are two Li--N--3H motifs centered at $0.25c$ and two centered at $0.75c$, each with a separation of $0.5b$.
These groupings of two motifs are separated from each other by an offset of $(0.5a,0.25b,0.5c)$.
As can be seen in Fig.~\ref{fig:struct}(a), Li--N--3H motifs that lie in the same $ab$-plane but do {\it not} share a H anion have alternating apical order; i.e., if the N anion is above Li in one motif, it will be below Li in the other.
The orthorhomic structure contains empty channels separating neighboring Li atoms along the $b$ axis, as shown schematically in Fig.~\ref{fig:struct}(c).

\begin{table*}
\setlength{\tabcolsep}{5pt}
\setlength{\extrarowheight}{4.5pt}
\caption{Calculated and Experimental Bulk Properties for Sr$_2$LiH$_2$N and La$_2$LiHN$_2$.}
\label{tab:bulk}
\centering
\begin{tabular}{lccccccc} \hline \hline
	\thead{Material} & \thead{Method} & \thead{$a$ ($\textrm{\AA}$)} & \thead{$b$ ($\textrm{\AA}$)} & \thead{$c$ ($\textrm{\AA}$)} & \thead{Volume (\AA$^3$)} & \thead{$E_{g,{\rm dir}}$ (eV)} & \thead{$E_{g,{\rm ind}}$ (eV)} \\ \hline
	Sr$_2$LiH$_2$N & Calc.& 3.79 & 7.23 & 14.69 & 402.81 & 2.49 & 2.40\\
			       & Ref.~\citenum{blaschkowski2007darstellung} & 3.70 & 7.47 & 13.30 & 367.60 &    &  \\
			       & Ref.~\citenum{liu2010synthesis} & 3.81 & 7.62 & 13.72 & 398.46 &   &  \\
	La$_2$LiHN$_2$ & Calc.& 3.84 & 7.24 & 13.08 & 363.38 & & \\ \hline \hline		 						
\end{tabular}
\end{table*}

\begin{figure}
\includegraphics[width=9cm]{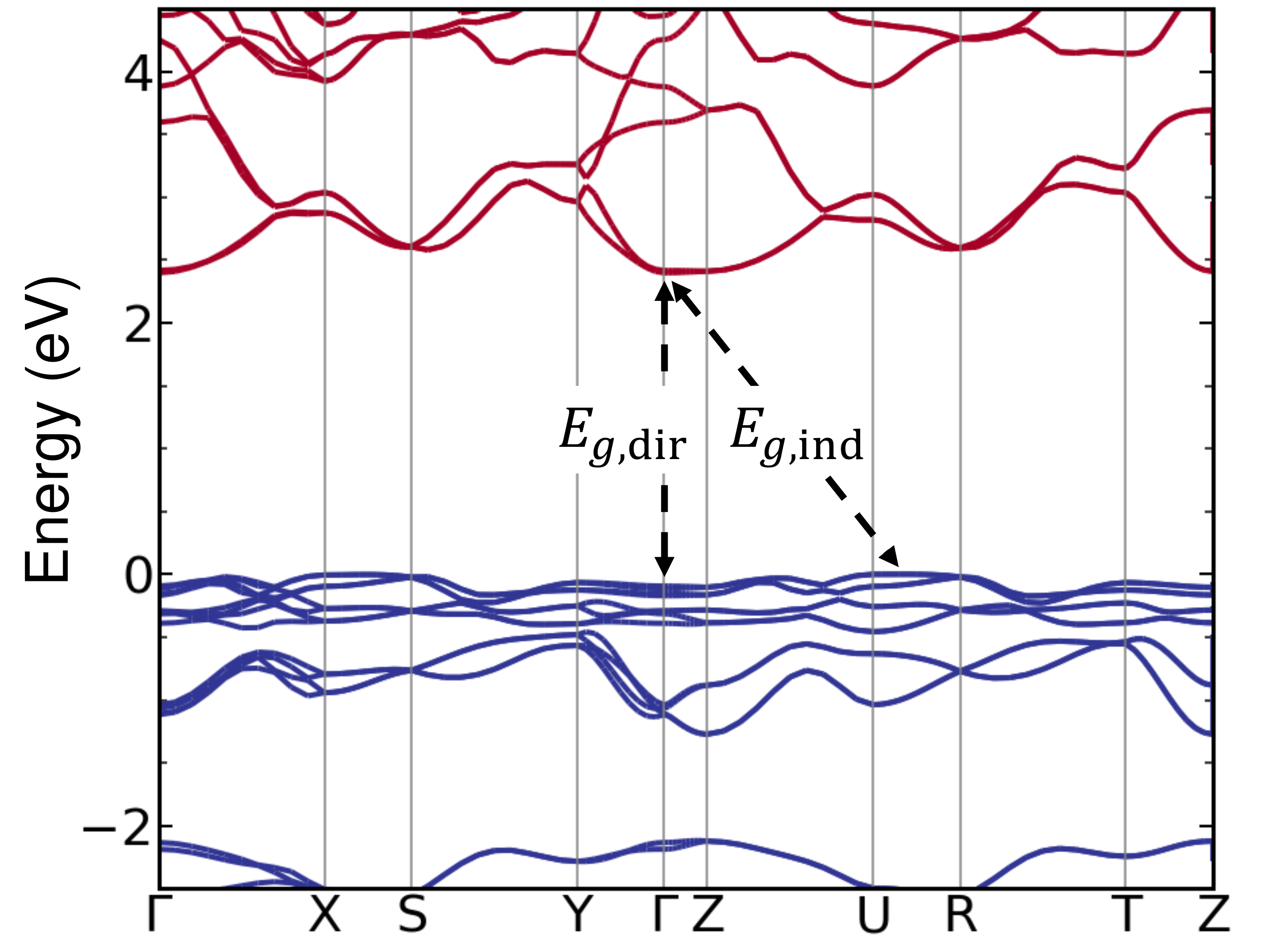}
\caption{The band structure of SLHN.}
\label{fig:bands}
\end{figure}

Our calculated band structure for SLHN is shown in Fig.~\ref{fig:bands}.
We find the band gap to be indirect, with a value of 2.40 eV.
The conduction-band minimum (CBM) is located at $\Gamma$, while the valence-band maximum (VBM) is located between the U and R points in the Brillouin zone.
The direct band gap at $\Gamma$ (2.49 eV) is only slightly larger.
The band gap is wide enough that suppressing electrical conductivity may be possible, which would enable SLHN to be used as an ionic electrolyte.

\subsection{Stability}
\label{sec:stab}

Our approach to evaluate stability begins with the thermodynamic stability condition:
\begin{equation}
	\Delta H^f(\rm{Sr}_2\rm{LiH}_2\rm{N}) = 2\Delta \mu_{\rm Sr}+\Delta \mu_{\rm Li} + 2\Delta \mu_{\rm H} + \Delta \mu_{\rm N} .
\label{eq:thermo}
\end{equation}
In this expression, $\Delta H^f$ refers to the enthalpy of formation, while the $\Delta \mu$ values are deviations from the chemical potential of the elemental phases for the four constituent elements.
For Sr, Li, and La, these phases are elemental metals, while for H and N, the reference phases are molecular H$_2$ and N$_2$,
all at $T=0$ K.
In order to prevent the formation of these phases, all $\Delta \mu$ values must be less than zero.
The total chemical potential for an element E is then given by:
\begin{equation}
	\mu_{\rm E}=E_{tot}({\rm E})+\Delta \mu_{\rm E} ,
\label{eq:chempot}
\end{equation}
where $E_{tot}({\rm E})$ is the total energy per atom of the elemental phase of E.

The existence of other compounds within the Sr-Li-H-N (or La-Li-H-N) phase space sets further limits on the values of $\Delta \mu$ for which SLHN (LLHN) can form.
These limits are quantified for some arbitrary compound Sr$_a$Li$_b$H$_c$N$_d$ as follows:
\begin{equation}
	a\Delta \mu_{\rm Sr}+b\Delta \mu_{\rm Li}+c\Delta \mu_{\rm H}+d\Delta \mu_{\rm N}\leq\Delta H^f({\rm{Sr}}_a{\rm{Li}}_b{\rm{H}}_c{\rm{N}}_d) .
\label{eq:limit}
\end{equation}
For each compound, these limits lead to equations with three unknowns, because the thermodynamic stability equation (eq~\ref{eq:thermo}) provides one constraint on the four variables.
In order to plot stability in two dimensions, we fix one of the remaining variables.
For our present purposes, we plot our results in the $\Delta \mu_{\rm Sr}$-vs.-$\Delta \mu_{\rm H}$ phase space, with $\Delta \mu_{\rm N}$ fixed.
Experimentally, $\Delta \mu_{\rm H}$ and $\Delta \mu_{\rm N}$ are the easiest of these four variables to control, as they correspond to partial pressures of H$_2$ and N$_2$.
For purposes of presenting our results, we identify regions of the stability diagrams corresponding to H-rich/H-poor and N-rich/N-poor conditions.
Immediately past the ``rich'' limit for an element, we expect another phase with a greater composition of that element to be thermodynamically favored over SLHN.

\begin{table}
\setlength{\tabcolsep}{5pt}
\setlength{\extrarowheight}{4pt}
\begin{tabular}{ccc} \hline \hline
\thead{Compound \\ } & \thead{$\Delta H^f$ (eV/f.u.) \\ (calc)} & \thead{$\Delta H^f$ (eV/f.u.) \\ (exp)} \\ \hline
Sr$_2$LiH$_2$N & --3.20 & -- \\
La$_2$LiHN$_2$ & --6.37 & -- \\
SrH$_2$ & --1.80 & --1.87\cite{lide2012crc} \\
LiH & --0.90 & --0.94\cite{lide2012crc} \\
Li$_3$N & --2.69 & --1.71\cite{o1975lithium} \\
LiSrH$_3$ & --1.10 & -- \\
Sr$_2$N & --2.59 & -- \\
Sr$_2$HN & --2.55 & -- \\\hline \hline
\end{tabular}
\caption{Calculated and Reported Enthalpies of Formation (in eV per Formula Unit) for Compounds Pertinent to this Study.}
\label{tab:enth}
\end{table}

\begin{figure*}
\includegraphics[width=18cm]{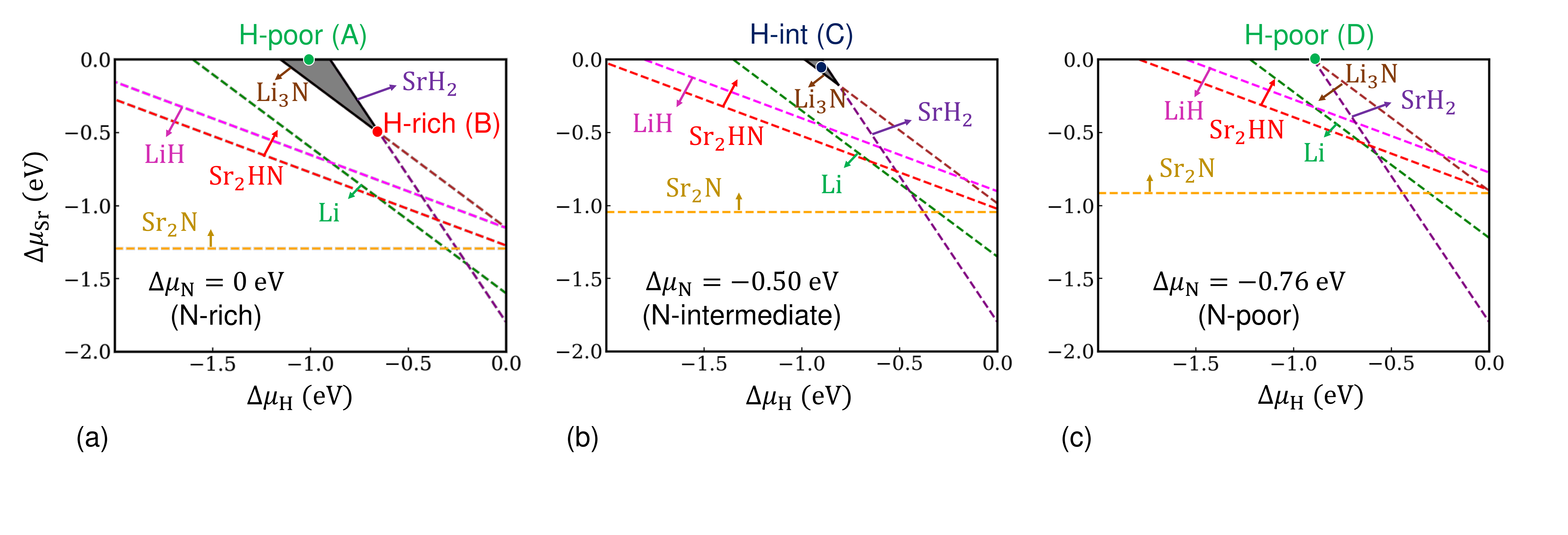}
\caption{
Phase diagram for Sr$_2$LiH$_2$N in $\Delta \mu_{\rm Sr}$-vs.-$\Delta \mu_{\rm H}$ phase space for three choices of $\Delta \mu_{\rm N}$. N-rich conditions are shown in (a), N-poor conditions in (c), and in (b), an intermediate choice of $\Delta \mu_{\rm N}$ is used. The dots indicate particular choices of chemical potentials, tabulated in Table~\ref{tab:chempot}, that are used to display the results of our defect calculations.
The region encompassing our chosen chemical potentials is shaded in gray.
In Fig.~\ref{fig:stability}(c), this region collapses to a single point.
Note that the conditions are labeled as (A), (B), (C), and (D) to correspond with the panels in Fig.~\ref{fig:defects}.}
\label{fig:stability}
\end{figure*}

In Fig.~\ref{fig:stability}, we plot the phase diagram for SLHN in the $\Delta \mu_{\rm Sr}$-vs.-$\Delta \mu_{\rm H}$ plane at three selected values of $\Delta \mu_{\rm N}$.
Our calculated enthalpies of formation for SLHN and its limiting phases are listed in Table~\ref{tab:enth}, alongside experimental values where available.
Our results show that, at $T$=0, SLHN is not stable---certain binary Sr--N compounds (such as Sr$_2$N) and also Sr$_2$HN will form preferentially at the chemical potentials where we would otherwise expect to observe the SLHN phase.
Given that experimental reports have not observed Sr$_2$HN or any binary Sr--N phases during synthesis of SLHN,\cite{blaschkowski2007darstellung,liu2010synthesis} we infer that careful synthesis techniques are able to create at least a metastable material.
We hypothesize that formation of other phases is kinetically suppressed, and/or the SLHN phase is entropically stabilized at finite temperature.
We identify a region of metastability, shaded in gray in Fig.~\ref{fig:stability}, where we will study the energetics of defect formation.
For the purposes of displaying the results of the defect calculations we will refer to conditions indicated by dots in Fig.~\ref{fig:stability}, for which we also tabulate values for $\Delta \mu$ in Table~\ref{tab:chempot}.

\begin{table*}
\setlength{\tabcolsep}{5pt}
\setlength{\extrarowheight}{4.5pt}
\caption{Selected Chemical Potentials, Labeled (A), (B), (C), and (D), Referring to Points Indicated in the Chemical Potential Stability Diagrams in Fig.~\ref{fig:stability}.}
\label{tab:chempot}
\centering
\begin{tabular}{lccccc} \hline \hline
	\thead{Condition} & \thead{$\Delta \mu_{\rm Sr}$ (eV)} & \thead{$\Delta \mu_{\rm Li}$ (eV)} & \thead{$\Delta \mu_{\rm H}$ (eV)} & \thead{$\Delta \mu_{\rm N}$ (eV)} & \thead{$\Delta \mu_{\rm O}$ (eV)}\\ \hline
	N-rich/H-poor (A) &   0    & --1.15 & --1.03 &   0    & --5.63\\
	N-rich/H-rich (B) & --0.51 & --0.90 & --0.65 &   0    & --5.13\\
	N-int/H-int (C)   & --0.09 & --0.78 & --0.88 & --0.50 & --5.54\\
	N-poor/H-poor (D) &   0    & --0.65 & --0.90 & --0.76 & --5.63\\ \hline \hline		 						
\end{tabular}
\end{table*}

We also calculated the stability phase space for LLHN, again considering several binary and ternary compounds.
Again, we found no chemical potentials for which LLHN will be stable.
The main culprit in limiting LLHN's stability is LaN, which is preferred to LLHN over a wide range of conditions.
Coupled with the lack of experimental evidence of its existence, it appears unlikely that LLHN can be stabilized.

\subsection{Defect Formation}
\label{sec:form}

\subsubsection{Native Point Defects}
\label{sec:native}

Point defects are ubiquitous in materials and play a critical role in determining their properties.
In the case of hydrogen electrolytes, the dominant charge carriers can be thought of as defects, be they hydrogen vacancies ($V_{\rm H}$) or hydrogen interstitials (H$_i$).
Vacancies are the predominant ionic charge carriers in hydrides and oxyhydrides,\cite{rowberg_AeH2_2018,liu2018formation} while interstitials are more important in proton conductors.\cite{rowberg_zirconates_2019,kreuer2003proton}
However, considering the ample amount of free space in the SLHN lattice, interstitials may well be important for hydride conduction in the nitride hydrides.

We calculate the formation energy of point defects using the equation:\cite{FreysoldtRMP2014}
\begin{multline}
	E^f(D^q) = E(D^q) - E_{\textrm{bulk}} + \\ \sum n_{\rm E}\mu_{\rm E} + qE_F + \Delta_{\textrm{corr}} ,
\label{eq:form}
\end{multline}
where $E^f(D^q)$ is the formation energy of defect $D$ in charge state $q$; $E(D^q)$ is the total energy of a supercell containing $D^q$; $E_{\rm bulk}$ is the total energy of the defect-free supercell; $n_{\rm E}$ is the number of atoms of species E added to or removed from the supercell, with $\mu_{\rm E}$ as defined in eq~\ref{eq:chempot}; $E_F$ is the Fermi level; and $\Delta_{\textrm{corr}}$ is a finite-size correction term.
Formation energy is related to the defect concentration by a Boltzmann relation:
\begin{equation}
	c = N_{\rm sites} \exp \left( -\frac{E^f}{k_BT}\right) \, ,
\label{eq:conc}
\end{equation}
where $N_\textrm{sites}$ represents the concentration of available defect sites.
Clearly, defect concentration increases exponentially as formation energy decreases, meaning that the most important defects will be those with the lowest formation energy at $E_F$.
Charge neutrality in the system then imposes a position of $E_F$ near the intersection of the lowest-energy positively and negatively charged defects.

\begin{figure*}
\includegraphics[width=18cm]{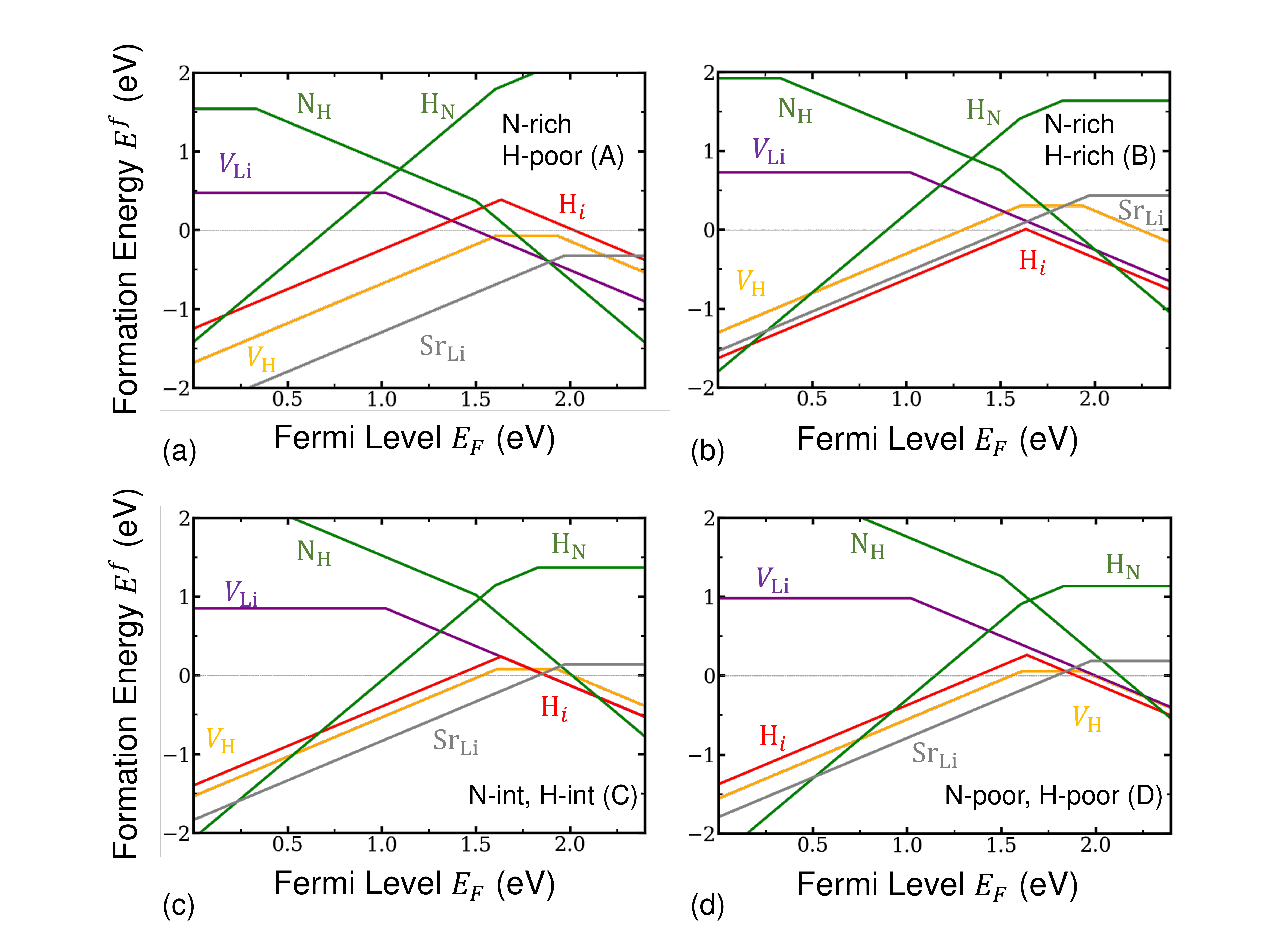}
\caption{Defect formation energies as a function of Fermi level in Sr$_2$LiH$_2$N under (a) N-rich/H-poor, (b) N-rich/H-rich, (c) N-intermediate/H-intermediate, and (d) N-poor/H-poor chemical potential conditions, as defined in Fig.~\ref{fig:stability} and Table~\ref{tab:chempot}. Labels (A), (B), (C), and (D) used here correspond to those used in Fig.~\ref{fig:stability} and Table~\ref{tab:chempot}.}
\label{fig:defects}
\end{figure*}

We plot the formation energy for various point defects as a function of $E_F$ over the band gap in Fig.~\ref{fig:defects} for the four chemical potential conditions listed in Table~\ref{tab:chempot}.
With regard to defects on H sites (vacancies and substitutional species), we have systematically investigated formation on apical sites (collinear with Li along the $c$ axis) and on sites along the $a$ axis relative to Li.
For all the defects we examined, the latter are lower in energy.
As for H$_i$, its location depends on the charge state: H$_i^-$ forms preferentially between two Li atoms along the $b$ axis, while H$_i^+$ bonds with a N anion.
As can be seen in Fig.~\ref{fig:defects}, the defect chemistry is highly dependent on the chemical potentials.

Under N-rich/H-poor conditions, shown in Fig.~\ref{fig:defects}(a), Sr$_{\rm Li}$, $V_{\rm Li}$, and N$_{{\rm H}}$ have negative formation energies at the Fermi level determined by charge neutrality, indicating that SLHN will be unstable.
However, at other conditions, the formation energy at which defect compensation occurs is positive.
One commonality of these other conditions is the ubiquity of H$_i^-$ as the dominant acceptor species.
Under N-rich/H-rich conditions [Fig.~\ref{fig:defects}(b)], H$_i^+$ will compensate H$_i^-$, implying that SLHN may allow for simultaneous stabilization of protons and hydride ions, which is a highly unusual but scientifically interesting possibility.\cite{norby2004hydrogen}
However, as conditions evolve toward the N-poor/H-poor limit, the cation antisite Sr$_{\rm Li}^+$ becomes the dominant donor defect.
At the N-poor/H-poor limit [Fig.~\ref{fig:defects}(d)], these species are accompanied by a large concentration of $V_{\rm H}^0$, which, while not influencing the location of $E_F$, may contribute to the movement of hydrogen.
In general, though, our observation of the favorability of H$_i$ supports the experimental observation of a significant amount of excess hydrogen content in SLHN reported by Liu $et$ $al.$\cite{liu2010synthesis}

\subsubsection{Oxygen Impurities}
\label{sec:oxy}

Both experimental reports on SLHN reported a high sensitivity to moisture.\cite{blaschkowski2007darstellung,liu2010synthesis}
We are therefore prompted to calculate the formation energy of several possible oxygen impurity configurations: substitutional O$_{\rm N}$, O$_{\rm H}$, and interstitial O$_i$.
Doing so will allow us to set some limits on how much oxygen in the environment can be tolerated.

\begin{figure}
\includegraphics[width=9cm]{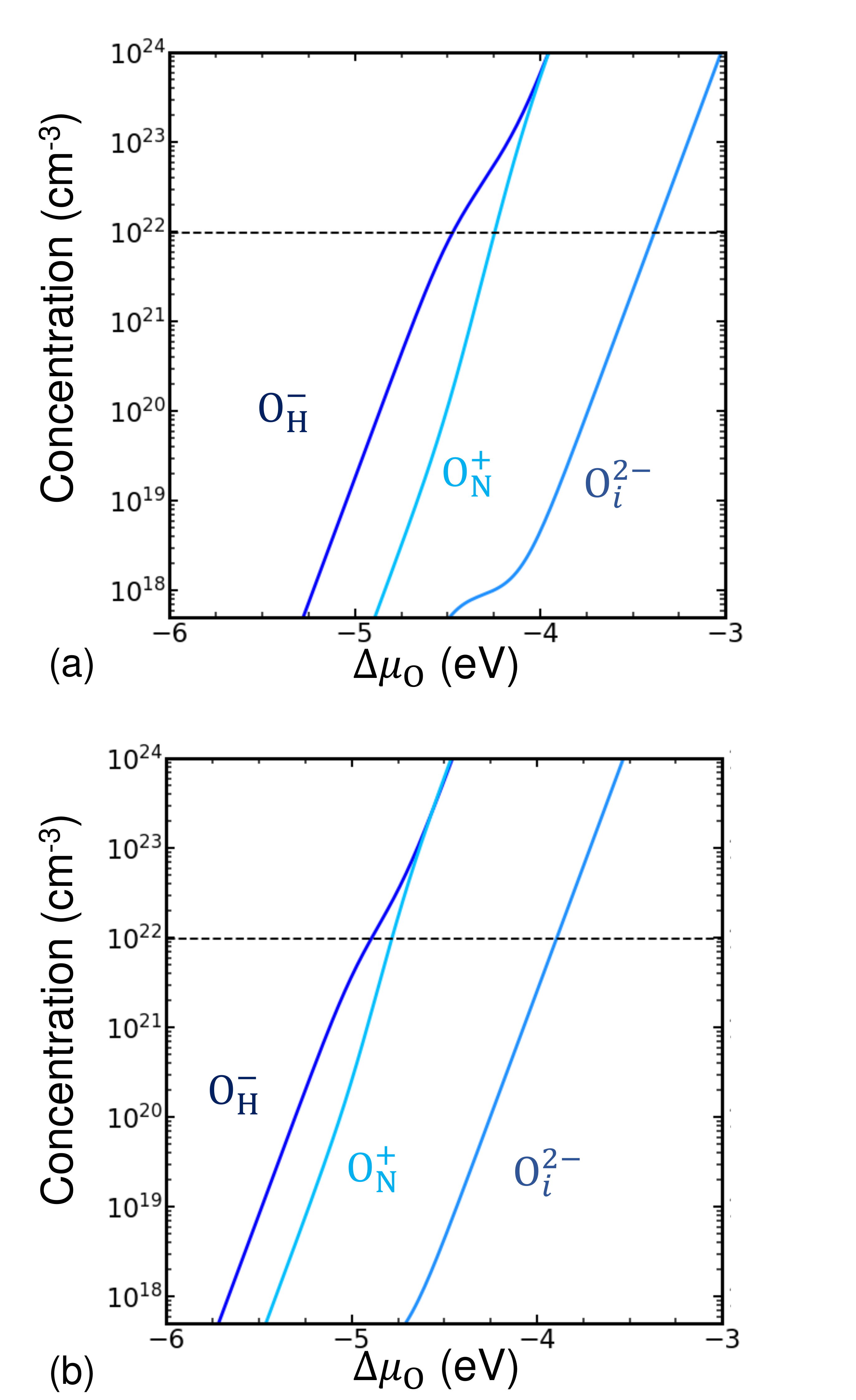}
\caption{Concentrations of oxygen impurities and H$_i^-$ in SLHN at $T=900$ K as a function of $\Delta\mu_{\rm O}$ for (a) N-rich/H-rich conditions (condition B in Table~\ref{tab:chempot}) and (b) N-poor/H-poor conditions (condition D in Table~\ref{tab:chempot}). The dashed horizontal line indicates defect concentrations at which the material will no longer be stable.}
\label{fig:o-defects}
\end{figure}

In Fig.~\ref{fig:o-defects}, we plot the concentration of oxygen impurities as a function of $\Delta\mu_{\rm O}$, which is correlated to experimental oxygen partial pressures.
We use $T=900$ K to generate the plot, as that corresponds to typical synthesis temperatures.\cite{blaschkowski2007darstellung}
We aim to identify oxygen chemical potential values for which the formation energy of oxygen defects remains positive; based on eq~\ref{eq:conc} this corresponds to an oxygen concentration just below 10$^{22}$ cm$^{-3}$, as indicated in Fig.~\ref{fig:o-defects} by a horizontal dashed line.
O$_{\rm H}^-$ and O$_{\rm N}^{+}$ turn out to be the dominant oxygen configurations; O$_i^{2-}$ is much less prevalent, in comparison.
Under N-rich/H-rich conditions [Fig.~\ref{fig:o-defects}(a)], oxygen impurities will cause the material to become unstable for $\Delta\mu_{\rm O}>-4.75$ eV.
We note that increasing concentrations of O$_{\rm H}^-$ will also cause the Fermi level to shift to lower energies, which suppresses the concentration of H$_i^-$ and limits the hydride conductivity.
The situation is worse for N-poor/H-poor conditions [Fig.~\ref{fig:o-defects}(b)], for which the limiting oxygen chemical potential already occurs at $\Delta\mu_{\rm O}=-5$ eV.
Thus, N-rich/H-rich conditions make the essential task of avoiding oxygen contamination more achievable.
Degradation upon oxygen exposure is common among many nitrides; fortunately, protective strategies such as encapsulation are available that have shown great promise.\cite{gregoire2008high}

\subsection{Hydride Migration}
\label{sec:mig}

\begin{figure*}
\includegraphics[width=18cm]{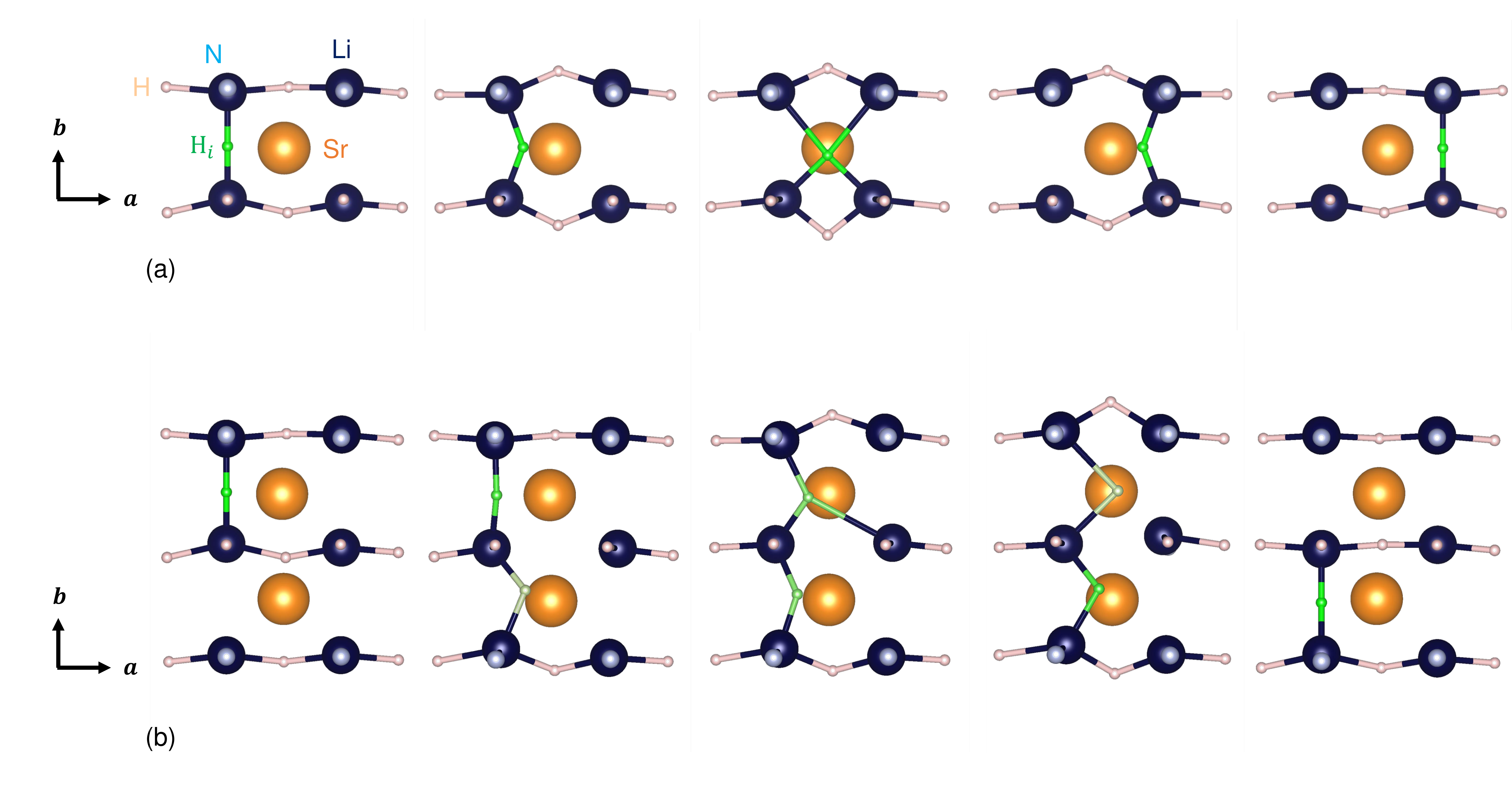}
\caption{Schematic depictions of hydride migration in Sr$_2$LiH$_2$N.
Mobile hydrogen species are colored green for clarity.
(a) Diffusion of H$_i^-$ down an empty channel along the $a$ axis. (b) Diffusion of H$_i^-$ along the $b$ axis.
}
\label{fig:neb}
\end{figure*}

The favorability of H$_i^-$ implies that it will be the most important defect for conductivity in SLHN.
Therefore, we consider its mobility using the NEB method.\cite{henkelman_neb}
With NEB, we can calculate the migration barrier $E_b$ for a particular path in the crystal, which is related to the activation energy $E_a$ for ionic conduction as:
\begin{equation}
	E_a=E_b+E^f ,
\label{eq:act}
\end{equation}
where $E^f$ is the formation energy.
For H$_i^-$, the formation energy ranges from approximately 0.01 eV under N-rich/H-rich conditions to 0.05 eV under N-poor/H-poor conditions---as a reminder, these are the chemical potential conditions under which SLHN remains stable with respect to native defects [as opposed to N-rich/H-poor conditions, Fig.~\ref{fig:defects}(a)].
These low formation energies imply that the migration barrier will be the limiting factor for ionic conduction.
A likely pathway for H$_i^-$ diffusion is migration down the $a$ axis in the empty channels separating chains of Li--N--3H motifs [see Fig.~\ref{fig:struct}(c)]; thus, we begin by calculating the energetic barrier to migration for this channel diffusion mechanism, shown schematically in Fig.~\ref{fig:neb}(a).
Our calculations reveal a symmetric barrier of 0.40 eV, which translates to an activation energy of 0.41--0.45 eV.

Another pathway for H$_i^-$ diffusion involves movement along the $b$ axis.
There are no free spaces for H$_i^-$ to move in this direction; however, diffusion may proceed via a ``kick-out'' process, whereby H$_i^-$ replaces a lattice hydride, which in turn becomes an H$_i^-$ in the neighboring channel.
This process is shown in Fig.~\ref{fig:neb}(b), wherein the initial and final states show H$_i^-$ in green, while intermediate images show the evolution between configurations by varying the color.
Diffusion along the $b$ axis has a higher migration barrier than does diffusion along $a$; the barrier of 0.59 eV indicates that this path will occur less frequently.

In contrast to the $a$- and $b$-directions, there are no obvious migration pathways along the $c$ axis.
It is likely that diffusion along the $c$ axis would require the passage of H$_i^-$ around relatively large Sr atoms.
We determined that the formation energy of H$_i^-$ in these regions is high, which means that the migration barrier will also be large.
Another option, based on movement of apical H anions, seems similarly unlikely, given that our formation energy calculations imply that there is limited tolerance for disorder on anionic apical sites in SLHN.
As a result, diffusion will be primarily anisotropic, and the lower energy pathway along the empty channels parallel to the $a$ axis should dominate hydride conduction in SLHN.
Its activation energy is lower than that measured in LaHO, which has demonstrated the highest measured solid-state hydride conductivity.\cite{fukui2019characteristic}

For comparison, we also consider hydride migration in LLHN.
Our NEB calculations show hydride migration to be considerably less favorable in LLHN.
H$_i^-$ diffusion along a pathway similar to the one shown in Fig.~\ref{fig:neb}(a), has a very high barrier, approximately 1.55 eV.
The size of the barrier appears to be connected to Li--Li spacing, which is larger in SLHN than in LLHN; as a result, diffusion of H$_i^-$ is more constricted in LLHN.
The large magnitudes of these migration barriers serve as a further indication that LLHN is not worth exploring as a hydride conductor.

\section{Conclusions}
\label{sec:conc}

In conclusion, we have calculated defect formation and migration properties in SLHN, a nitride hydride with intriguing prospects as a solid-state hydride electrolyte.
We also considered the La-based analogue to SLHN; however, its poor hydride mobility renders it significantly less attractive.
For SLHN, we have identified chemical potential conditions under which defect formation energies are positive, implying that careful synthesis can lead to a metastable material.
Defects will be quite prevalent in SLHN, particularly H$_i^-$, which will compensate with H$_i^+$ under N-rich conditions or Sr$_{\rm Li}^+$ under more N-poor conditions.
The low formation energy of H$_i^-$, coupled with its low migration barrier of 0.40 eV along the $a$ axis, leads us to predict an activation energy that can be as low as 0.4 eV, which is superior to the measured activation energy in the most conductive oxyhydrides.
While oxygen incorporation can be damaging for SLHN's stability and conductivity, our results demonstrate that, with careful synthesis and usage, its performance will rival that of the best solid-state hydrogen electrolytes currently known.

\section*{Acknowledgements}

A.J.E.R. was supported by the National Science Foundation (NSF) Graduate Research Fellowship Program under Grant No. 1650114. Any opinions, findings, and conclusions or recommendations expressed in this material are those of the author(s) and do not necessarily reflect the views of the NSF.
C.G.V.d.W. was supported by the Office of Science of the U.S. Department of Energy (DOE) (Grant No. DE-FG02-07ER46434).
We acknowledge the use of the Center for Scientific Computing supported by the California NanoSystems Institute and the Materials Research Science and Engineering Center (MRSEC) at UC Santa Barbara through NSF DMR 1720256 and NSF CNS 0960316. We also acknowledge computational resources provided through the National Energy Research Scientific Computing Center, a DOE Office of Science User Facility supported by the Office of Science of the U.S. DOE under Contract No. DE-AC02-05CH11231.

\bibliography{Nitride-hydrides}

\providecommand{\latin}[1]{#1}
\makeatletter
\providecommand{\doi}
  {\begingroup\let\do\@makeother\dospecials
  \catcode`\{=1 \catcode`\}=2 \doi@aux}
\providecommand{\doi@aux}[1]{\endgroup\texttt{#1}}
\makeatother
\providecommand*\mcitethebibliography{\thebibliography}
\csname @ifundefined\endcsname{endmcitethebibliography}
  {\let\endmcitethebibliography\endthebibliography}{}
\begin{mcitethebibliography}{46}
\providecommand*\natexlab[1]{#1}
\providecommand*\mciteSetBstSublistMode[1]{}
\providecommand*\mciteSetBstMaxWidthForm[2]{}
\providecommand*\mciteBstWouldAddEndPuncttrue
  {\def\EndOfBibitem{\unskip.}}
\providecommand*\mciteBstWouldAddEndPunctfalse
  {\let\EndOfBibitem\relax}
\providecommand*\mciteSetBstMidEndSepPunct[3]{}
\providecommand*\mciteSetBstSublistLabelBeginEnd[3]{}
\providecommand*\EndOfBibitem{}
\mciteSetBstSublistMode{f}
\mciteSetBstMaxWidthForm{subitem}{(\alph{mcitesubitemcount})}
\mciteSetBstSublistLabelBeginEnd
  {\mcitemaxwidthsubitemform\space}
  {\relax}
  {\relax}

\bibitem[Schlapbach and Z\"{u}ttel(2001)Schlapbach, and
  Z\"{u}ttel]{schlapbach_hydrogen-storage_2001}
Schlapbach,~L.; Z\"{u}ttel,~A. Hydrogen-storage materials for mobile
  applications. \emph{Nature} \textbf{2001}, \emph{414}, 353--358\relax
\mciteBstWouldAddEndPuncttrue
\mciteSetBstMidEndSepPunct{\mcitedefaultmidpunct}
{\mcitedefaultendpunct}{\mcitedefaultseppunct}\relax
\EndOfBibitem
\bibitem[Steele and Heinzel(2001)Steele, and Heinzel]{SteeleNAT2001}
Steele,~B.~C.; Heinzel,~A. Materials for fuel-cell technologies. \emph{Nature}
  \textbf{2001}, \emph{414}, 345--352\relax
\mciteBstWouldAddEndPuncttrue
\mciteSetBstMidEndSepPunct{\mcitedefaultmidpunct}
{\mcitedefaultendpunct}{\mcitedefaultseppunct}\relax
\EndOfBibitem
\bibitem[Crabtree and Dresselhaus(2008)Crabtree, and
  Dresselhaus]{crabtree2008hydrogen}
Crabtree,~G.~W.; Dresselhaus,~M.~S. The hydrogen fuel alternative. \emph{MRS
  Bull.} \textbf{2008}, \emph{33}, 421--428\relax
\mciteBstWouldAddEndPuncttrue
\mciteSetBstMidEndSepPunct{\mcitedefaultmidpunct}
{\mcitedefaultendpunct}{\mcitedefaultseppunct}\relax
\EndOfBibitem
\bibitem[Agency(2019)]{IEA2019}
Agency,~I.~E. \emph{The Future of Hydrogen: Seizing today's opportunities};
  International Energy Agency: Paris, FR, 2019\relax
\mciteBstWouldAddEndPuncttrue
\mciteSetBstMidEndSepPunct{\mcitedefaultmidpunct}
{\mcitedefaultendpunct}{\mcitedefaultseppunct}\relax
\EndOfBibitem
\bibitem[Rikukawa and Sanui(2000)Rikukawa, and Sanui]{rikukawa2000proton}
Rikukawa,~M.; Sanui,~K. Proton-conducting polymer electrolyte membranes based
  on hydrocarbon polymers. \emph{Prog. Polymer Sci.} \textbf{2000}, \emph{25},
  1463--1502\relax
\mciteBstWouldAddEndPuncttrue
\mciteSetBstMidEndSepPunct{\mcitedefaultmidpunct}
{\mcitedefaultendpunct}{\mcitedefaultseppunct}\relax
\EndOfBibitem
\bibitem[Wang \latin{et~al.}(2011)Wang, Chen, Mishler, Cho, and
  Adroher]{wang2011review}
Wang,~Y.; Chen,~K.~S.; Mishler,~J.; Cho,~S.~C.; Adroher,~X.~C. A review of
  polymer electrolyte membrane fuel cells: technology, applications, and needs
  on fundamental research. \emph{Appl. Ener.} \textbf{2011}, \emph{88},
  981--1007\relax
\mciteBstWouldAddEndPuncttrue
\mciteSetBstMidEndSepPunct{\mcitedefaultmidpunct}
{\mcitedefaultendpunct}{\mcitedefaultseppunct}\relax
\EndOfBibitem
\bibitem[Jena(2011)]{jena2011materials}
Jena,~P. Materials for hydrogen storage: past, present, and future. \emph{J.
  Phys. Chem. Lett.} \textbf{2011}, \emph{2}, 206--211\relax
\mciteBstWouldAddEndPuncttrue
\mciteSetBstMidEndSepPunct{\mcitedefaultmidpunct}
{\mcitedefaultendpunct}{\mcitedefaultseppunct}\relax
\EndOfBibitem
\bibitem[Kreuer(2003)]{kreuer2003proton}
Kreuer,~K.-D. Proton-conducting oxides. \emph{Annu. Rev. Mater. Res.}
  \textbf{2003}, \emph{33}, 333--359\relax
\mciteBstWouldAddEndPuncttrue
\mciteSetBstMidEndSepPunct{\mcitedefaultmidpunct}
{\mcitedefaultendpunct}{\mcitedefaultseppunct}\relax
\EndOfBibitem
\bibitem[Iwahara \latin{et~al.}(1993)Iwahara, Yajima, Hibino, Ozaki, and
  Suzuki]{iwahara_zirc_1993}
Iwahara,~H.; Yajima,~T.; Hibino,~T.; Ozaki,~K.; Suzuki,~H. Protonic conduction
  in calcium, strontium and barium zirconates. \emph{Solid State Ionics}
  \textbf{1993}, \emph{61}, 65--69\relax
\mciteBstWouldAddEndPuncttrue
\mciteSetBstMidEndSepPunct{\mcitedefaultmidpunct}
{\mcitedefaultendpunct}{\mcitedefaultseppunct}\relax
\EndOfBibitem
\bibitem[Stambouli and Traversa(2002)Stambouli, and
  Traversa]{StambouliRSER2002}
Stambouli,~A.~B.; Traversa,~E. Solid oxide fuel cells (SOFCs): a review of an
  environmentally clean and efficient source of energy. \emph{Renew. Sustain.
  Ener. Rev.} \textbf{2002}, \emph{6}, 433--455\relax
\mciteBstWouldAddEndPuncttrue
\mciteSetBstMidEndSepPunct{\mcitedefaultmidpunct}
{\mcitedefaultendpunct}{\mcitedefaultseppunct}\relax
\EndOfBibitem
\bibitem[Verbraeken \latin{et~al.}(2014)Verbraeken, Cheung, Suard, and
  Irvine]{verbraeken_high_2014}
Verbraeken,~M.~C.; Cheung,~C.; Suard,~E.; Irvine,~J. T.~S. High H$^-$ ionic
  conductivity in barium hydride. \emph{Nat. Mater.} \textbf{2014}, \emph{14},
  95--100\relax
\mciteBstWouldAddEndPuncttrue
\mciteSetBstMidEndSepPunct{\mcitedefaultmidpunct}
{\mcitedefaultendpunct}{\mcitedefaultseppunct}\relax
\EndOfBibitem
\bibitem[Kobayashi \latin{et~al.}(2016)Kobayashi, Hinuma, Matsuoka, Watanabe,
  Iqbal, Hirayama, Yonemura, Kamiyama, Tanaka, and Kanno]{kobayashi2016pure}
Kobayashi,~G.; Hinuma,~Y.; Matsuoka,~S.; Watanabe,~A.; Iqbal,~M.; Hirayama,~M.;
  Yonemura,~M.; Kamiyama,~T.; Tanaka,~I.; Kanno,~R. Pure {H}$^-$ conduction in
  oxyhydrides. \emph{Science} \textbf{2016}, \emph{351}, 1314--1317\relax
\mciteBstWouldAddEndPuncttrue
\mciteSetBstMidEndSepPunct{\mcitedefaultmidpunct}
{\mcitedefaultendpunct}{\mcitedefaultseppunct}\relax
\EndOfBibitem
\bibitem[Kobayashi \latin{et~al.}(2012)Kobayashi, Hernandez, Sakaguchi, Yajima,
  Roisnel, Tsujimoto, Morita, Noda, Mogami, Kitada, Ohkura, Hosokawa, Li,
  Hayashi, Kusano, Kim, Tsuji, Fujiwara, Matsushita, Yoshimura, Takegoshi,
  Inoue, Takano, and Kageyama]{kobayashi_oxyhydride_2012}
Kobayashi,~Y. \latin{et~al.}  An oxyhydride of {BaTiO}$_3$ exhibiting hydride
  exchange and electronic conductivity. \emph{Nat. Mater.} \textbf{2012},
  \emph{11}, 507--511\relax
\mciteBstWouldAddEndPuncttrue
\mciteSetBstMidEndSepPunct{\mcitedefaultmidpunct}
{\mcitedefaultendpunct}{\mcitedefaultseppunct}\relax
\EndOfBibitem
\bibitem[Bridges \latin{et~al.}(2006)Bridges, Fernandez-Alonso, Goff, and
  Rosseinsky]{bridges2006observation}
Bridges,~C.~A.; Fernandez-Alonso,~F.; Goff,~J.~P.; Rosseinsky,~M.~J.
  Observation of Hydride Mobility in the Transition-Metal Oxide Hydride
  $\textrm{La} \textrm{Sr} \textrm{Co} \textrm{O}_3 \textrm{H}_{0.7}$.
  \emph{Adv. Mater.} \textbf{2006}, \emph{18}, 3304--3308\relax
\mciteBstWouldAddEndPuncttrue
\mciteSetBstMidEndSepPunct{\mcitedefaultmidpunct}
{\mcitedefaultendpunct}{\mcitedefaultseppunct}\relax
\EndOfBibitem
\bibitem[Rowberg \latin{et~al.}(2018)Rowberg, Weston, and Van~de
  Walle]{rowberg_AeH2_2018}
Rowberg,~A. J.~E.; Weston,~L.; Van~de Walle,~C.~G. Ion-transport engineering of
  alkaline-earth hydrides for hydride electrolyte applications. \emph{Chem.
  Mater.} \textbf{2018}, \emph{30}, 5878--5885\relax
\mciteBstWouldAddEndPuncttrue
\mciteSetBstMidEndSepPunct{\mcitedefaultmidpunct}
{\mcitedefaultendpunct}{\mcitedefaultseppunct}\relax
\EndOfBibitem
\bibitem[Bai \latin{et~al.}(2018)Bai, He, Zhu, and Mo]{bai2018first}
Bai,~Q.; He,~X.; Zhu,~Y.; Mo,~Y. First-Principles Study of Oxyhydride H--Ion
  Conductors: Toward Facile Anion Conduction in Oxide-Based Materials.
  \emph{ACS Appl. Ener. Mater.} \textbf{2018}, \emph{1}, 1626--1634\relax
\mciteBstWouldAddEndPuncttrue
\mciteSetBstMidEndSepPunct{\mcitedefaultmidpunct}
{\mcitedefaultendpunct}{\mcitedefaultseppunct}\relax
\EndOfBibitem
\bibitem[Liu \latin{et~al.}(2018)Liu, Bj{\o}rheim, and
  Haugsrud]{liu2018formation}
Liu,~X.; Bj{\o}rheim,~T.~S.; Haugsrud,~R. Formation of defects and their
  effects on hydride ion transport properties in a series of K$_2$NiF$_4$-type
  oxyhydrides. \emph{J. Mater. Chem. A} \textbf{2018}, \emph{6},
  1454--1461\relax
\mciteBstWouldAddEndPuncttrue
\mciteSetBstMidEndSepPunct{\mcitedefaultmidpunct}
{\mcitedefaultendpunct}{\mcitedefaultseppunct}\relax
\EndOfBibitem
\bibitem[Kageyama \latin{et~al.}(2018)Kageyama, Hayashi, Maeda, Attfield,
  Hiroi, Rondinelli, and Poeppelmeier]{kageyama2018expanding}
Kageyama,~H.; Hayashi,~K.; Maeda,~K.; Attfield,~J.~P.; Hiroi,~Z.;
  Rondinelli,~J.~M.; Poeppelmeier,~K.~R. Expanding frontiers in materials
  chemistry and physics with multiple anions. \emph{Nat. Commun.}
  \textbf{2018}, \emph{9}, 772\relax
\mciteBstWouldAddEndPuncttrue
\mciteSetBstMidEndSepPunct{\mcitedefaultmidpunct}
{\mcitedefaultendpunct}{\mcitedefaultseppunct}\relax
\EndOfBibitem
\bibitem[Aleksanyan and Dolukhanyan(2001)Aleksanyan, and
  Dolukhanyan]{aleksanyan2001combustion}
Aleksanyan,~A.; Dolukhanyan,~S. Combustion of niobium in hydrogen and nitrogen.
  Synthesis of niobium hydrides and hydridonitrides. \emph{Int. J. Hydrog.
  Ener.} \textbf{2001}, \emph{26}, 429--433\relax
\mciteBstWouldAddEndPuncttrue
\mciteSetBstMidEndSepPunct{\mcitedefaultmidpunct}
{\mcitedefaultendpunct}{\mcitedefaultseppunct}\relax
\EndOfBibitem
\bibitem[Chen \latin{et~al.}(2002)Chen, Xiong, Luo, Lin, and
  Tan]{chen2002interaction}
Chen,~P.; Xiong,~Z.; Luo,~J.; Lin,~J.; Tan,~K.~L. Interaction of hydrogen with
  metal nitrides and imides. \emph{Nature} \textbf{2002}, \emph{420},
  302--304\relax
\mciteBstWouldAddEndPuncttrue
\mciteSetBstMidEndSepPunct{\mcitedefaultmidpunct}
{\mcitedefaultendpunct}{\mcitedefaultseppunct}\relax
\EndOfBibitem
\bibitem[Tapia-Ruiz \latin{et~al.}(2013)Tapia-Ruiz, Sorbie, Vach{\'e}, Hoang,
  and Gregory]{tapia2013rapid}
Tapia-Ruiz,~N.; Sorbie,~N.; Vach{\'e},~N.; Hoang,~T.~K.; Gregory,~D.~H. Rapid
  microwave synthesis, characterization and reactivity of lithium nitride
  hydride, Li$_4$NH. \emph{Materials} \textbf{2013}, \emph{6}, 5410--5426\relax
\mciteBstWouldAddEndPuncttrue
\mciteSetBstMidEndSepPunct{\mcitedefaultmidpunct}
{\mcitedefaultendpunct}{\mcitedefaultseppunct}\relax
\EndOfBibitem
\bibitem[Reckeweg and DiSalvo(2002)Reckeweg, and DiSalvo]{reckeweg2002alkaline}
Reckeweg,~O.; DiSalvo,~F.~J. Alkaline earth metal nitride compounds with the
  composition $M_2NX$ ($M$= Ca, Sr, Ba; $X$= , H, Cl or Br). \emph{Solid state
  sciences} \textbf{2002}, \emph{4}, 575--584\relax
\mciteBstWouldAddEndPuncttrue
\mciteSetBstMidEndSepPunct{\mcitedefaultmidpunct}
{\mcitedefaultendpunct}{\mcitedefaultseppunct}\relax
\EndOfBibitem
\bibitem[Chemnitzer \latin{et~al.}(2005)Chemnitzer, Auffermann, T{\"o}bbens,
  and Kniep]{chemnitzer2005sr2n}
Chemnitzer,~R.; Auffermann,~G.; T{\"o}bbens,~D.~M.; Kniep,~R. (Sr$_2$N) H:
  Untersuchungen zur Redox-Intercalation von Wasserstoff in Sr$_2$N. \emph{Z.
  Anorg. Allg. Chem.} \textbf{2005}, \emph{631}, 1813--1817\relax
\mciteBstWouldAddEndPuncttrue
\mciteSetBstMidEndSepPunct{\mcitedefaultmidpunct}
{\mcitedefaultendpunct}{\mcitedefaultseppunct}\relax
\EndOfBibitem
\bibitem[Wegner \latin{et~al.}(1992)Wegner, Essmann, Bock, Jacobs, and
  Fischer]{wegner1992structure}
Wegner,~B.; Essmann,~R.; Bock,~J.; Jacobs,~H.; Fischer,~P. Structure and
  H--ionic-conductivity of barium hydride nitride, Ba$2$H(D)N. \emph{Eur. J.
  Solid State Inorg. Chem.} \textbf{1992}, \emph{29}, 1217--1227\relax
\mciteBstWouldAddEndPuncttrue
\mciteSetBstMidEndSepPunct{\mcitedefaultmidpunct}
{\mcitedefaultendpunct}{\mcitedefaultseppunct}\relax
\EndOfBibitem
\bibitem[Altorfer \latin{et~al.}(1994)Altorfer, B{\"u}hrer, Winkler, Coddens,
  Essmann, and Jacobs]{altorfer1994Ba2NH}
Altorfer,~F.; B{\"u}hrer,~W.; Winkler,~B.; Coddens,~G.; Essmann,~R.; Jacobs,~H.
  H$^-$-jump diffusion in barium-nitride-hydride Ba$_2$NH. \emph{Solid State
  Ionics} \textbf{1994}, \emph{70}, 272--277\relax
\mciteBstWouldAddEndPuncttrue
\mciteSetBstMidEndSepPunct{\mcitedefaultmidpunct}
{\mcitedefaultendpunct}{\mcitedefaultseppunct}\relax
\EndOfBibitem
\bibitem[Blaschkowski and Schleid(2007)Blaschkowski, and
  Schleid]{blaschkowski2007darstellung}
Blaschkowski,~B.; Schleid,~T. Darstellung und Kristallstruktur des
  Lithium-Strontium-Hydridnitrids LiSr$_2$H$_2$N. \emph{Z. Anorg. Allg. Chem.}
  \textbf{2007}, \emph{633}, 2644--2648\relax
\mciteBstWouldAddEndPuncttrue
\mciteSetBstMidEndSepPunct{\mcitedefaultmidpunct}
{\mcitedefaultendpunct}{\mcitedefaultseppunct}\relax
\EndOfBibitem
\bibitem[Liu \latin{et~al.}(2010)Liu, Liu, Si, and Zhang]{liu2010synthesis}
Liu,~D.; Liu,~Q.; Si,~T.; Zhang,~Q. Synthesis and crystal structure of a novel
  nitride hydride Sr$_2$LiNH$_2$. \emph{J. Alloys Cmpnds.} \textbf{2010},
  \emph{495}, 272--274\relax
\mciteBstWouldAddEndPuncttrue
\mciteSetBstMidEndSepPunct{\mcitedefaultmidpunct}
{\mcitedefaultendpunct}{\mcitedefaultseppunct}\relax
\EndOfBibitem
\bibitem[Matsui \latin{et~al.}(2019)Matsui, Kobayashi, Suzuki, Watanabe,
  Kubota, Iwasaki, Yonemura, Hirayama, and Kanno]{matsui2019ambient}
Matsui,~N.; Kobayashi,~G.; Suzuki,~K.; Watanabe,~A.; Kubota,~A.; Iwasaki,~Y.;
  Yonemura,~M.; Hirayama,~M.; Kanno,~R. Ambient pressure synthesis of
  La$_2$LiHO$_3$ as a solid electrolyte for a hydrogen electrochemical cell.
  \emph{J. Am. Ceram. Soc.} \textbf{2019}, \emph{102}, 3228--3235\relax
\mciteBstWouldAddEndPuncttrue
\mciteSetBstMidEndSepPunct{\mcitedefaultmidpunct}
{\mcitedefaultendpunct}{\mcitedefaultseppunct}\relax
\EndOfBibitem
\bibitem[Rowberg \latin{et~al.}(2021)Rowberg, Weston, and Van~de
  Walle]{rowberg_oxyhydrides_2020}
Rowberg,~A. J.~E.; Weston,~L.; Van~de Walle,~C.~G. Defect chemistry and
  hydrogen transport in La/Sr-based oxyhydrides. \emph{J. Phys. Chem. C}
  \textbf{2021}, \emph{125}, 2250\relax
\mciteBstWouldAddEndPuncttrue
\mciteSetBstMidEndSepPunct{\mcitedefaultmidpunct}
{\mcitedefaultendpunct}{\mcitedefaultseppunct}\relax
\EndOfBibitem
\bibitem[Kohn and Sham(1965)Kohn, and Sham]{kohn_self-consistent_1965}
Kohn,~W.; Sham,~L.~J. Self-Consistent Equations Including Exchange and
  Correlation Effects. \emph{Phys. Rev.} \textbf{1965}, \emph{140},
  A1133--A1138\relax
\mciteBstWouldAddEndPuncttrue
\mciteSetBstMidEndSepPunct{\mcitedefaultmidpunct}
{\mcitedefaultendpunct}{\mcitedefaultseppunct}\relax
\EndOfBibitem
\bibitem[Hohenberg and Kohn(1964)Hohenberg, and Kohn]{hohenberg1964}
Hohenberg,~P.; Kohn,~W. \emph{Phys. Rev.} \textbf{1964}, \emph{136}, B864\relax
\mciteBstWouldAddEndPuncttrue
\mciteSetBstMidEndSepPunct{\mcitedefaultmidpunct}
{\mcitedefaultendpunct}{\mcitedefaultseppunct}\relax
\EndOfBibitem
\bibitem[Heyd \latin{et~al.}(2003)Heyd, Scuseria, and Ernzerhof]{heyd_hse}
Heyd,~J.; Scuseria,~G.~E.; Ernzerhof,~M. Hybrid functionals based on a screened
  Coulomb potential. \emph{J. Chem. Phys.} \textbf{2003}, \emph{118},
  8207--8215\relax
\mciteBstWouldAddEndPuncttrue
\mciteSetBstMidEndSepPunct{\mcitedefaultmidpunct}
{\mcitedefaultendpunct}{\mcitedefaultseppunct}\relax
\EndOfBibitem
\bibitem[Henkelman and J\'{o}nsson(2000)Henkelman, and
  J\'{o}nsson]{henkelman_neb}
Henkelman,~G.; J\'{o}nsson,~H. Improved tangent estimate in the nudged elastic
  band method for finding minimum energy paths and saddle points. \emph{J.
  Chem. Phys.} \textbf{2000}, \emph{113}, 9978--9985\relax
\mciteBstWouldAddEndPuncttrue
\mciteSetBstMidEndSepPunct{\mcitedefaultmidpunct}
{\mcitedefaultendpunct}{\mcitedefaultseppunct}\relax
\EndOfBibitem
\bibitem[Fukui \latin{et~al.}(2019)Fukui, Iimura, Tada, Fujitsu, Sasase,
  Tamatsukuri, Honda, Ikeda, Otomo, and Hosono]{fukui2019characteristic}
Fukui,~K.; Iimura,~S.; Tada,~T.; Fujitsu,~S.; Sasase,~M.; Tamatsukuri,~H.;
  Honda,~T.; Ikeda,~K.; Otomo,~T.; Hosono,~H. Characteristic fast H- ion
  conduction in oxygen-substituted lanthanum hydride. \emph{Nat. Commun.}
  \textbf{2019}, \emph{10}, 1--8\relax
\mciteBstWouldAddEndPuncttrue
\mciteSetBstMidEndSepPunct{\mcitedefaultmidpunct}
{\mcitedefaultendpunct}{\mcitedefaultseppunct}\relax
\EndOfBibitem
\bibitem[Kresse and Furthm\"{u}ller(1996)Kresse, and
  Furthm\"{u}ller]{kresse_vasp}
Kresse,~G.; Furthm\"{u}ller,~J. Efficient iterative schemes for ab initio
  total-energy calculations using a plane-wave basis set. \emph{Phys. Rev. B}
  \textbf{1996}, \emph{54}, 11169\relax
\mciteBstWouldAddEndPuncttrue
\mciteSetBstMidEndSepPunct{\mcitedefaultmidpunct}
{\mcitedefaultendpunct}{\mcitedefaultseppunct}\relax
\EndOfBibitem
\bibitem[Bl\"{o}chl(1994)]{blochl_paw1}
Bl\"{o}chl,~P.~E. Projector augmented-wave method. \emph{Phys. Rev. B}
  \textbf{1994}, \emph{50}, 17953--17979\relax
\mciteBstWouldAddEndPuncttrue
\mciteSetBstMidEndSepPunct{\mcitedefaultmidpunct}
{\mcitedefaultendpunct}{\mcitedefaultseppunct}\relax
\EndOfBibitem
\bibitem[Kresse and Joubert(1999)Kresse, and Joubert]{kresse_paw2}
Kresse,~G.; Joubert,~D. From ultrasoft pseudopotentials to the projector
  augmented-wave method. \emph{Phys. Rev. B} \textbf{1999}, \emph{59},
  1758--1775\relax
\mciteBstWouldAddEndPuncttrue
\mciteSetBstMidEndSepPunct{\mcitedefaultmidpunct}
{\mcitedefaultendpunct}{\mcitedefaultseppunct}\relax
\EndOfBibitem
\bibitem[Perdew \latin{et~al.}(1996)Perdew, Burke, and
  Ernzerhof]{perdew1996generalized}
Perdew,~J.~P.; Burke,~K.; Ernzerhof,~M. Generalized gradient approximation made
  simple. \emph{Phys. Rev. Lett.} \textbf{1996}, \emph{77}, 3865\relax
\mciteBstWouldAddEndPuncttrue
\mciteSetBstMidEndSepPunct{\mcitedefaultmidpunct}
{\mcitedefaultendpunct}{\mcitedefaultseppunct}\relax
\EndOfBibitem
\bibitem[Momma and Izumi(2011)Momma, and Izumi]{momma2011vesta}
Momma,~K.; Izumi,~F. VESTA3 for three-dimensional visualization of crystal,
  volumetric and morphology data. \emph{J. Appl. Crystal.} \textbf{2011},
  \emph{44}, 1272--1276\relax
\mciteBstWouldAddEndPuncttrue
\mciteSetBstMidEndSepPunct{\mcitedefaultmidpunct}
{\mcitedefaultendpunct}{\mcitedefaultseppunct}\relax
\EndOfBibitem
\bibitem[Lide(2012)]{lide2012crc}
Lide,~D.~R. \emph{CRC Handbook of Chemistry and Physics}; 2012; pp 5--5 --
  5--60\relax
\mciteBstWouldAddEndPuncttrue
\mciteSetBstMidEndSepPunct{\mcitedefaultmidpunct}
{\mcitedefaultendpunct}{\mcitedefaultseppunct}\relax
\EndOfBibitem
\bibitem[O'Hare and Johnson(1975)O'Hare, and Johnson]{o1975lithium}
O'Hare,~P.; Johnson,~G.~K. Lithium nitride (Li$_3$N): standard enthalpy of
  formation by solution calorimetry. \emph{J. Chem. Thermodyn.} \textbf{1975},
  \emph{7}, 13--20\relax
\mciteBstWouldAddEndPuncttrue
\mciteSetBstMidEndSepPunct{\mcitedefaultmidpunct}
{\mcitedefaultendpunct}{\mcitedefaultseppunct}\relax
\EndOfBibitem
\bibitem[Rowberg \latin{et~al.}(2019)Rowberg, Weston, and Van~de
  Walle]{rowberg_zirconates_2019}
Rowberg,~A. J.~E.; Weston,~L.; Van~de Walle,~C.~G. Optimizing proton
  conductivity in zirconates through defect engineering. \emph{ACS Appl. Ener.
  Mater.} \textbf{2019}, \emph{2}, 2611--2619\relax
\mciteBstWouldAddEndPuncttrue
\mciteSetBstMidEndSepPunct{\mcitedefaultmidpunct}
{\mcitedefaultendpunct}{\mcitedefaultseppunct}\relax
\EndOfBibitem
\bibitem[Freysoldt \latin{et~al.}(2014)Freysoldt, Grabowski, Hickel,
  Neugebauer, Kresse, Janotti, and Van~de Walle]{FreysoldtRMP2014}
Freysoldt,~C.; Grabowski,~B.; Hickel,~T.; Neugebauer,~J.; Kresse,~G.;
  Janotti,~A.; Van~de Walle,~C.~G. First-principles calculations for point
  defects in solids. \emph{Rev. Mod. Phys.} \textbf{2014}, \emph{86}, 253\relax
\mciteBstWouldAddEndPuncttrue
\mciteSetBstMidEndSepPunct{\mcitedefaultmidpunct}
{\mcitedefaultendpunct}{\mcitedefaultseppunct}\relax
\EndOfBibitem
\bibitem[Norby \latin{et~al.}(2004)Norby, Wider{\o}e, Gl{\"o}ckner, and
  Larring]{norby2004hydrogen}
Norby,~T.; Wider{\o}e,~M.; Gl{\"o}ckner,~R.; Larring,~Y. Hydrogen in oxides.
  \emph{Dalton Trans.} \textbf{2004}, 3012--3018\relax
\mciteBstWouldAddEndPuncttrue
\mciteSetBstMidEndSepPunct{\mcitedefaultmidpunct}
{\mcitedefaultendpunct}{\mcitedefaultseppunct}\relax
\EndOfBibitem
\bibitem[Gregoire \latin{et~al.}(2008)Gregoire, Kirby, Scopelianos, Lee, and
  van Dover]{gregoire2008high}
Gregoire,~J.~M.; Kirby,~S.~D.; Scopelianos,~G.~E.; Lee,~F.~H.; van Dover,~R.~B.
  High mobility single crystalline ScN and single-orientation epitaxial YN on
  sapphire via magnetron sputtering. \emph{J. Appl. Phys.} \textbf{2008},
  \emph{104}, 074913\relax
\mciteBstWouldAddEndPuncttrue
\mciteSetBstMidEndSepPunct{\mcitedefaultmidpunct}
{\mcitedefaultendpunct}{\mcitedefaultseppunct}\relax
\EndOfBibitem
\end{mcitethebibliography}

\end{document}